\documentclass[12pt]{article} % Example of a standard class

\usepackage[a4paper]{geometry}
\usepackage[utf8]{inputenc}
\usepackage{graphicx}
\usepackage{dcolumn}
\usepackage{bm}
\usepackage{verbatim}
\usepackage{soul}
\usepackage[dvipsnames]{xcolor}
\colorlet{RED}{red}
\usepackage{braket}
\usepackage{dsfont}
\usepackage{amsmath,float}
\usepackage{url}
\usepackage{ctable}
\usepackage{soul}
\usepackage[USenglish]{babel}%naustrian
\usepackage{titlesec}
\usepackage{gensymb}
\usepackage{subcaption}
\usepackage{authblk}

\usepackage{xr}
\usepackage{hyperref}

\titleformat{\section}{\normalfont\large\bfseries}{\thesection}{10pt}{}
\titleformat{\subsection}{\normalfont}{\thesubsection}{10pt}{\ul}
\titlespacing{\subsection}{0pt}{10pt}{4pt}
\titleformat{\subsubsection}{\normalfont\large}{\thesubsection}{10pt}{}
\titlespacing{\subsubsection}{0pt}{\parskip}{4pt}

\begin{document}

\title{\textbf{Non-degenerate SPDC photon-pair source for UV-A illumination}}
\author[1,2]{Preetisha Goswami}
\author[2,*]{Marta Gilaberte Basset}
\author[3,4]{Jorge Fuenzalida}
\author[3]{Markus Gräfe}
\author[2]{Valerio Flavio Gili}

\affil[1]{\textit{Friedrich-Schiller-University Jena, Institute of Applied Physics, Abbe Centre of Photonics, Albert-Einstein-Str. 6, 07745, Jena, Germany.}}
\affil[2]{\textit{Fraunhofer Institute for Applied Optics and Precision Engineering IOF, Albert-Einstein-Str. 7, 07745, Jena, Germany.}}
\affil[3]{\textit{Institute for Applied Physics, Technical University of Darmstadt, Otto-Berndt-Straße 3, 64287 Darmstadt, Germany.}}
\affil[4]{\textit{Present address: ICFO-Institut de Ciencies Fotoniques, The Barcelona Institute of Science and Technology, 08860 Castelldefels (Barcelona), Spain.}}
\affil[*]{\textit{marta.gilaberte.basset@iof.fraunhofer.de}}

\date{} % Optional: You can add a date here or leave it empty

\maketitle

\newpage
%---------------------ABSTRACT-----------------------
\section*{Abstract}

We present a frequency-correlated non-degenerate photon-pair source consisting of a second-order nonlinear crystal that generates ultraviolet UV-A and infrared light via spontaneous parametric down-conversion. Quantum imaging and sensing techniques like quantum imaging with undetected light and quantum ghost imaging leverage on wavelength correlations between down-converted photon pairs to decouple sensing and detection wavelengths, thereby exploiting established camera technology within the visible spectrum. %spectral regions like the VIS.
Our results open up novel quantum sensing application scenarios in the ultraviolet domain, with potential implications for advancements in biomedical and non-destructive testing fields.

\newpage
%%%%%%%%%%%%%%%%%%%%%%%%%%  body  %%%%%%%%%%%%%%%%%%%%%%%%%%
\section{Introduction}
 Quantum imaging and sensing techniques have the potential to surpass several classical limitations, such as resolution beyond the Abbe limit and noise reduction below the shot-noise limit \cite{MGB19, PAM19, MG16}. Here we highlight a different approach to the establishment of a "quantum advantage", namely the possibility of illuminating an object with a different wavelength than the one used for detection, thus compensating for limitations of the existing camera technology in specific wavelength ranges, especially when low light levels are required \cite{IK20, PAM15}. The most widespread effect used to generate non-classical light is via spontaneous parametric down-conversion (SPDC), where a pump photon interacts with a second-order nonlinear crystal and is converted into two lower energy correlated photons~\cite{Walborn2010}. In particular, frequency correlations are a key element for quantum imaging and sensing techniques, where phase-matching in nonlinear crystals can be engineered to control the signal and idler emission wavelengths. Therefore, the interest in developing a wider variety of SPDC biphoton sources has grown in recent years, especially towards the near- \cite{MH24} and mid-infrared \cite{MK21}. Most of the experimental demonstrations of imaging and spectroscopy techniques have therefore focused on the infrared (IR) or terahertz (THz) range of the electromagnetic spectrum \cite{DK16,CL2022,VFG22,VFG23,MK20}. Although many applications require IR or THz illumination \cite{MALI18,MAM18,BR00,TAG15}, others would benefit from ultraviolet (UV) illumination in quantum imaging and spectroscopy techniques \cite{LUN05,ALM16,FGA14}. Specifically for the quantum imaging with undetected light (QIUL) technique, the spatial resolution of exploiting the transversal momentum correlations between the down-converted photons is determined by the illumination wavelength \cite{JFU22}. This feature is particularly attractive when applied to UV imaging, as the resolution would be bound by the UV photon wavelength, while the twin IR photon would be used for the spatially resolved detection.
 Up to date, quantum state generation in the UV range is still challenging. Single-photon sources, have been demonstrated in quantum dots \cite{SCH24} and 2D-materials \cite{BOU16}. Conversely, photon-pair sources are more attractive in quantum imaging and spectroscopy applications, due to the possibility to exploit inherent frequency correlations of the generated photons to realise so-called "two-color" schemes: one photon (the idler) is typically generated in a wavelength range where camera detection technology is underdeveloped and interacts with the sample of interest, while the twin photon (the signal) is typically generated in the visible range (VIS), to exploit cheap, established and low noise silicon cameras. Two-color schemes have been either based on the ghost imaging (GI) technique, where a single-pixel detector (the bucket detector) is needed for the idler photon, or on QIUL, where no bucket detector for the idler photon is needed. As a consequence, two-color quantum GI has mainly been demonstrated in the near-infrared \cite{ASP15,VFG23}, whereas imaging/spectroscopy with undetected light has been demonstrated in the mid-infrared \cite{IK20, DK16} and in the terahertz \cite{MK20}. Despite the established implementation of such two-color schemes, photon-pair sources in which one of the twin photons is in the UV range remain more elusive, and no two-color quantum sensing schemes in the UV has been so far implemented. Only one non-degenerate photon-pair source for UV illumination has been reported, based on the third-order nonlinear process of spontaneous four-wave mixing (SFWM) \cite{LOH23}. The authors use a xenon gas-filled fiber to generate photon pairs, requiring pressure control on the gas for wavelength tunability. In this work, we exploit the birefringent phase matching on a Beta Barium Borate (BBO) crystal to generate correlated biphotons via SPDC. The signal light ranges between 532 nm and 1108 nm, while the idler light wavelength is angle-tunable in the range between 347 nm and 532 nm. Hence, the generated photon pair span from the UV-A range to the VIS, and the NIR, covering a large portion of the electromagnetic spectrum of high interest for biomedical applications.

%%%%%%%%%%%%%%%%%%%%%%%%%%  Experimental setup and results  %%%%%%%%%%%%%%%%%%%%%%%%%%
\section{SPDC source characterization}
\begin{figure}[htbp]
\centering\includegraphics[width=0.9\textwidth]{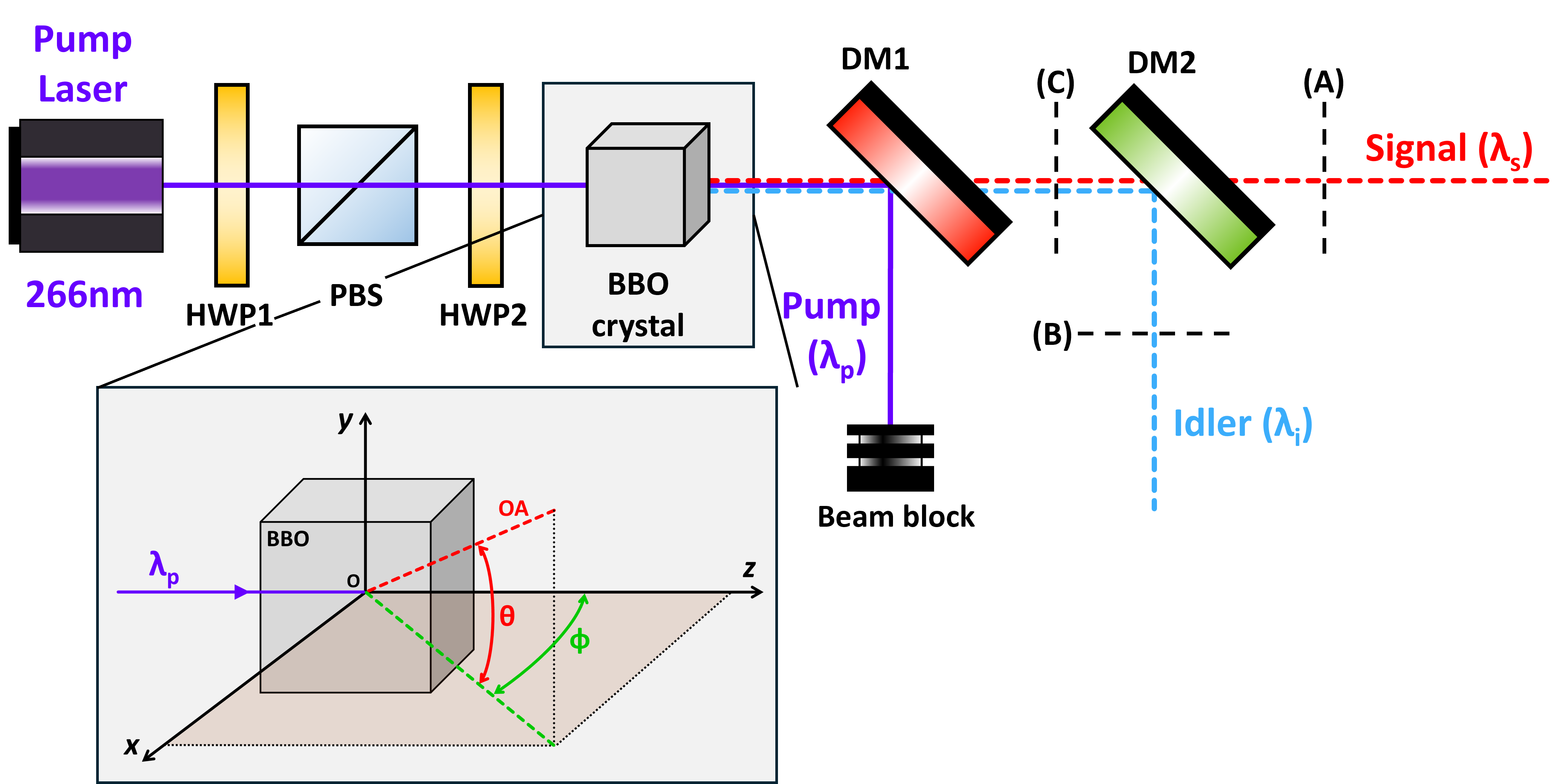}
\caption{Experimental setup for the photon-pair source characterization. The power of the pump laser is controlled with a half-waveplate (HWP1) and a polarization beam splitter (PBS), while a second HWP (HWP2) adjusts pump polarization for phase-matching. The BBO crystal is mounted on a five-axis stage that allows for its careful alignment to match the angular phase-matching conditions. Two dichroic mirrors (DM1 and DM2) separate the signal, idler, and pump light into different paths. The black dashed lines indicate the positions where the measuring devices where placed for the different analysis performed: light was fiber-coupled into a spectrometer at (C) to measure the down-converted spectra, and light was fiber-coupled into single-photon detectors (SPDs) at (A) and (B) to collect the signal and idler light, respectively. %, and an EMCCD camera was placed at (B) to estimate the UV photon-flux.
The inset shows the orientation of the BBO crystal and its OA (highlighted in red) with respect to the incoming pump beam, which is represented along the z-axis. The angles $\theta$ and $\phi$ define the rotations exploited during the SPDC spectral characterization. The initial orientation is $\theta_\mathrm{0}=42.7\degree$ and $\phi_\mathrm{0}=0$. For that position, the projection of the OA (highlighted in green) is in the pump beam direction.}
\label{fig:setup}
\end{figure}

The source we present makes use of a $2~\mathrm{mm}$ long BBO crystal with a $6x6~\mathrm{mm}^2$ aperture to generate down-converted pairs of correlated photons (signal and idler) through a birefringent type-I SPDC process. The nonlinear crystal is pumped with a $266~\mathrm{nm}$ CW laser with $200~\mathrm{mW}$ output power and a $400~\mathrm{\mu m}$ beam waist. As energy and momentum must be conserved through the generation process, the signal and idler photons show spectral and spatial correlations. In this work, we focus on the spectral correlations and we first analyze the down-converted spectral range accessible through angle tuning of the nonlinear crystal. Secondly, we verify the generation of photon pairs in the spontaneous regime by measuring the change in the detected single-photon and coincidence rates when varying the pump power and estimate the pair generation rate of the source. %Finally, we estimate the UV photon flux generated by analysing the image of the UV beam generated with an electron multiplying charge coupled device (EMCCD) camera.
A sketch of the experimental setup for the source characterization is presented in Fig.~\ref{fig:setup}. 

\subsection{Down-converted spectra}
First, the birefringence of the BBO is exploited to study the available range of down-converted spectra. The free space signal and idler beams are fiber-coupled into an Ocean Optics QE Pro spectrometer at position (C) (see Fig.~\ref{fig:setup}). %with a coupling efficiency of \mgb{XX\%}.
The initial angular position of the BBO is set to the position where the crystal surface is perpendicular to the incoming pump beam. At this initial position, the optical axis (OA) of the BBO crystal is at $\phi_\mathrm{0}=0\degree$ and $\theta_\mathrm{0}=42.7\degree$ from the direction of the incoming pump beam (see Fig.~\ref{fig:setup}). The phase-matching condition at this orientation of the BBO generates down-converted photon pairs at $904~\mathrm{nm}$ and $377~\mathrm{nm}$.
%The BBO crystal is cut to ensure that when it is perpendicular to the pump beam, the phase matching angles $\phi_\mathrm{PM}=90\degree$ and $\theta_\mathrm{PM}=42.7\degree$ (see Fig.~\ref{fig:setup}) generate down-converted photon pairs at $910~\mathrm{nm}$ and $376~\mathrm{nm}$. 
Different angular positions for $\phi$ and $\theta$ are carefully scanned relative to this initial orientation ($\phi_\mathrm{0}, \theta_\mathrm{0}$) to study the spectral range of down-converted light accessible with the BBO crystal.

In uniaxial anisotropic crystals, e.g. BBO crystals, the angle between the OA and the direction of the pump ($\theta^\prime$) defines the value of the extraordinary refractive index of the crystal following~\cite{Boyd}:

\begin{align}\label{Eq:BBOellipsoid}
   \frac{1}{n_\mathrm{e}(\theta^\prime)^2} =\frac{\sin^2 \theta^\prime}{\tilde{n}_\mathrm{e}^2} + \frac{\cos^2 \theta^\prime}{n_\mathrm{o}^2},
\end{align}

where $n_\mathrm{e}(\theta^\prime)$ is the refractive index of the extraordinary wave at the effective angle $\theta^\prime$, $\tilde{n}_\mathrm{e}$ represents the principal value of the extraordinary refractive index and $n_\mathrm{o}$ is the refractive index along the ordinary axis. The effective angle $\theta^\prime$ is a combination of $\theta$ and $\phi$ (Fig.~\ref{fig:setup}).

%The ordinary refractive index of the crystal does not depend on the angle $\theta^\prime$.

Although the largest effect on the extraordinary refractive index is expected when varying $\theta$, the adjustment of $\phi$ also induces a variation. This can be understood from the expression

\begin{align}\label{Eq:effective_angle}
  \theta^\prime=\arccos\{\cos(\theta)\cos(\phi)\},
\end{align}

and the reference values $\theta_\mathrm{0}=42.7\degree$ and $\phi_\mathrm{0}=0\degree$, which imply a stronger effect of $\theta$ than $\phi$ on $\theta^\prime$. The expression for the effective angle $\theta^\prime$ can be obtained by applying the corresponding rotation matrix relative to the initial direction of the OA. The rotation matrix depends on the specific order of the applied rotations. A rotation first around z-, then around y-, and finally around x-axis is described by

\begin{align}\label{Eq:Rxyz}
R_{XYZ} =\notag
\\
& \hspace{-1.3cm}
{\small
\begin{pmatrix}
\cos\gamma \cos\phi & -\sin\gamma \cos\theta + \cos\gamma \sin\phi \sin\theta & \sin\gamma \sin\theta + \cos\gamma \sin\phi \cos\theta \\
\sin\gamma \cos\phi & \cos\gamma \cos\theta + \sin\gamma \sin\phi \sin\theta  & -\cos\gamma \sin\theta + \sin\gamma \sin\phi \cos\theta \\
-\sin\phi           & \cos\phi \sin\theta                                     & \cos\phi \cos\theta
\end{pmatrix}.
}
\end{align}

However, an exchange of order between the rotations around the x- and y-axes results in a different combined rotation matrix $R_{YXZ}$ (Eq.~\ref{Eq:Ryxz}):

\begin{align}\label{Eq:Ryxz}
R_{YXZ} = \notag
\\
& \hspace{-1cm}
{\small
\begin{pmatrix}
\cos\gamma \cos\phi - \sin\gamma \sin\theta \sin\phi & -\sin\gamma \cos\theta & \cos\gamma \sin\phi + \sin\gamma \sin\theta \cos\phi \\
\sin\gamma \cos\phi + \cos\gamma \sin\theta \sin\phi & \cos\gamma \cos\theta & \sin\gamma \sin\phi - \cos\gamma \sin\theta \cos\phi \\
-\cos\theta \sin\phi & \sin\theta & \cos\theta \cos\phi
\end{pmatrix}.
}
\end{align}

In both performed rotations, the term describing the final orientation of the optical axis relative to the pump beam (along the z-axis; see Fig.~\ref{fig:setup}) remains unchanged. Consequently, the effective angle ($\theta^\prime$) is given by Eq.~\ref{Eq:effective_angle}.

%Anyway, the term that describes the final rotation of the OA from the pump beam direction (along the z-axis; see Fig.~\ref{fig:setup}) is the same for both performed rotations. Therefore, the effective angle $\theta^\prime$ follows Eq.~\ref{Eq:effective_angle}. %\mgb{See Supplementary material for the derivation of the effective angle expression?}

The angle around the z-axis, $\gamma$ in Eq.~\ref{Eq:Rxyz} and Eq.~\ref{Eq:Ryxz}, does not affect either of the refractive indices since the index ellipsoid defining the refractive indices is symmetric around the OA for uniaxial crystals \cite{Boyd}. Therefore, it is kept fixed for all measurements presented in this work.
%\mgb{ask Preetisha for data relative to this position, I think some graphs need a re-naming of x-axis--> Yes, and also the previous statement needs to be motivated (graph?)}.
\begin{figure}[h]
    \centering
    \begin{subfigure}{0.45\textwidth}
        \includegraphics[width=\textwidth]{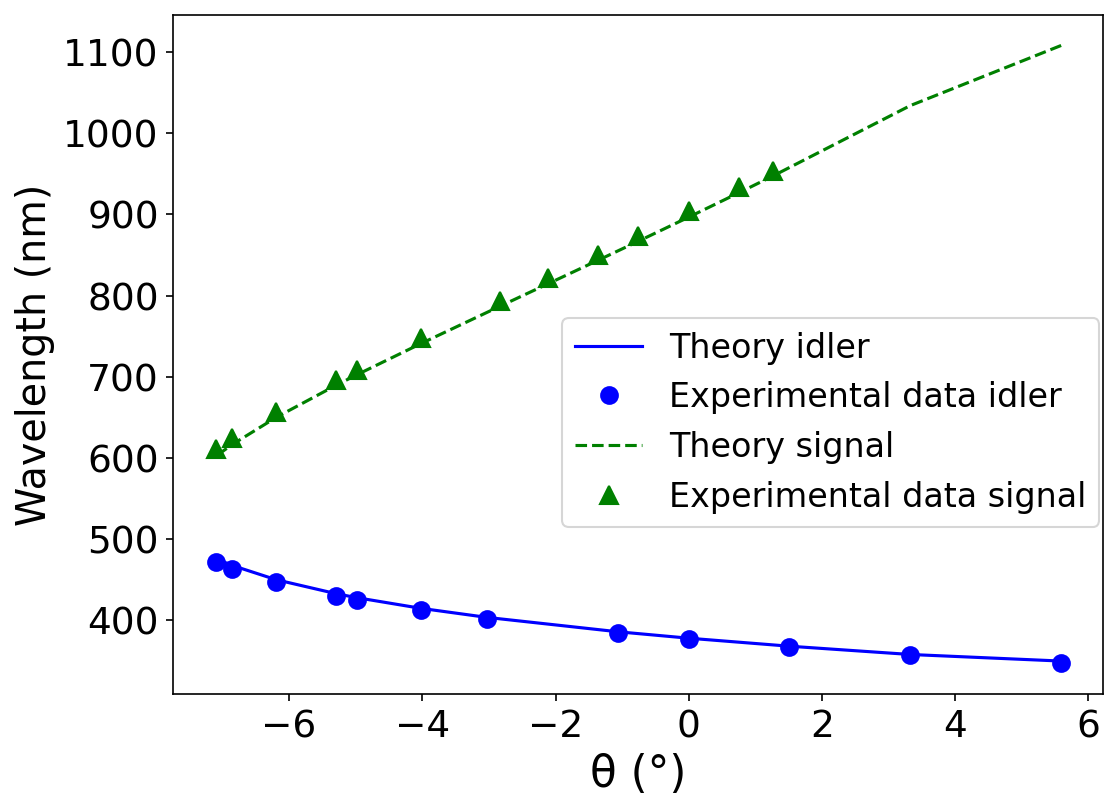}
        \caption{}
        \label{fig:SPDCspectra_a}
    \end{subfigure}
		\hfill
    \begin{subfigure}{0.45\textwidth}
        \includegraphics[width=\textwidth]{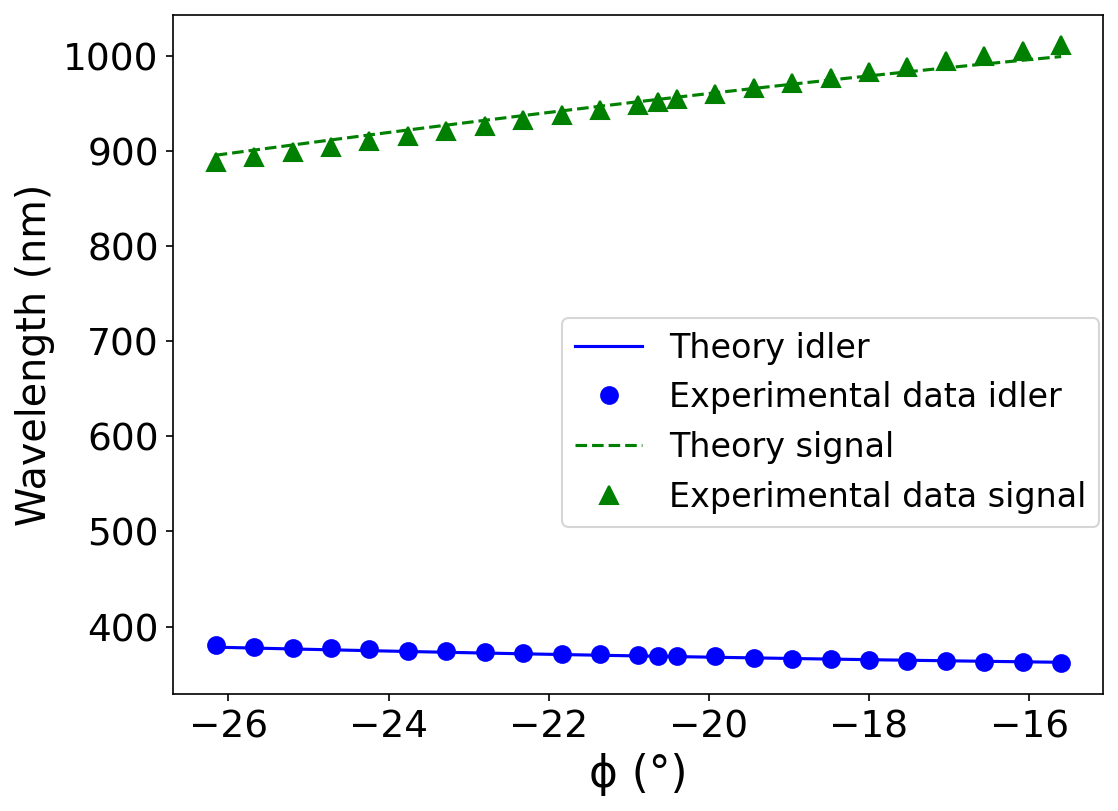}
        \caption{}
        \label{fig:SPDCspectra_b}
    \end{subfigure}
    \caption{a) Non-degenerate spectral range achieved by varying $\theta$ and mantaining $\phi$ fixed at $\phi_0$. The rotation matrix used to determine the effective angle $\theta^\prime$ for calculating the predicted spectra in this scenario is shown in Eq.~\ref{Eq:Rxyz}. Note that the signal wavelength is recorded over a shorter angular range compared to the idler wavelength data, due to the decreased detection efficiency of our spectrometer for longer wavelengths. Signal wavelengths are nonetheless calculated and plotted following the SPDC energy conservation formula up to 1108.1 nm, corresponding to the shortest idler we measure at 471.42 nm. Additionally, this graph shows data collected across three separate measurements due to experimental reasons. In the first measurement, both signal and idler wavelengths are recorded simultaneously in plane A (see Fig~\ref{fig:setup}) for angular variations up to $-4\degree$. In the second and third measurements, signal and idler are detected separately for angular tilts above $-4\degree$ using planes A and B, respectively. Consequently, for these angular values, the experimental data of the photon pairs appear mismatched. b) Non-degenerate spectral range achieved by varying $\phi$ and mantaining $\theta$ fixed at $\theta_0-2.5\degree$. The rotation matrix used to determine the effective angle $\theta^\prime$ for calculating the predicted spectra in this scenario is shown in Eq.~\ref{Eq:Ryxz}.}
    \label{fig:SPDCspectra}
\end{figure}

The non-degenerate wavelength range accessible by angle tuning is shown in Fig.~\ref{fig:SPDCspectra}. From the data shown in Fig.~\ref{fig:SPDCspectra_a}, signal wavelength can be tuned over a $\mathrm{496.9~nm}$ range (from $\mathrm{611.20~nm}$ to $\mathrm{1108.1~nm}$) when varying the crystal orientation by $\theta=12.69\degree$, even though the longest wavelengths are not observed due to experimental reasons. This corresponds to a wavelength tuning rate of $\mathrm{0.0255~\degree/nm}$. In comparison, the idler wavelength exhibits a tuning rate of $\mathrm{0.1023~\degree/nm}$ (covering $\mathrm{347.39~nm}$ to $\mathrm{471.42~nm}$), indicating that it is less sensitive to angular changes.
Figure~\ref{fig:SPDCspectra_b} shows the accessible wavelength range when $\phi$ is used for angular tuning. The corresponding spectra are more robust against angular variations around the y-axis. Consequently, a broader angular scan around $\phi$ produces a narrower spectral range than that achieved when varying $\theta$. These findings align with the theoretical expectations described previously. Furthermore, other regions of the spectra can be achieved by careful alignment of the BBO orientation. For example, degenerate emission at $\mathrm{532~nm}$ can also be achieved by carefully orienting the BBO crystal to $\phi=10.7\degree$ and adjusting $\theta$ by $\sim4\degree$. The total range observed under different crystal orientations spand from 532~nm to 1108~nm for the signal wavelength, and from 532~nm to 347~nm for the idler wavelength.

\subsection{Photon-pair generation}
\begin{figure}[htbp]
\centering\includegraphics[width=0.5\textwidth]{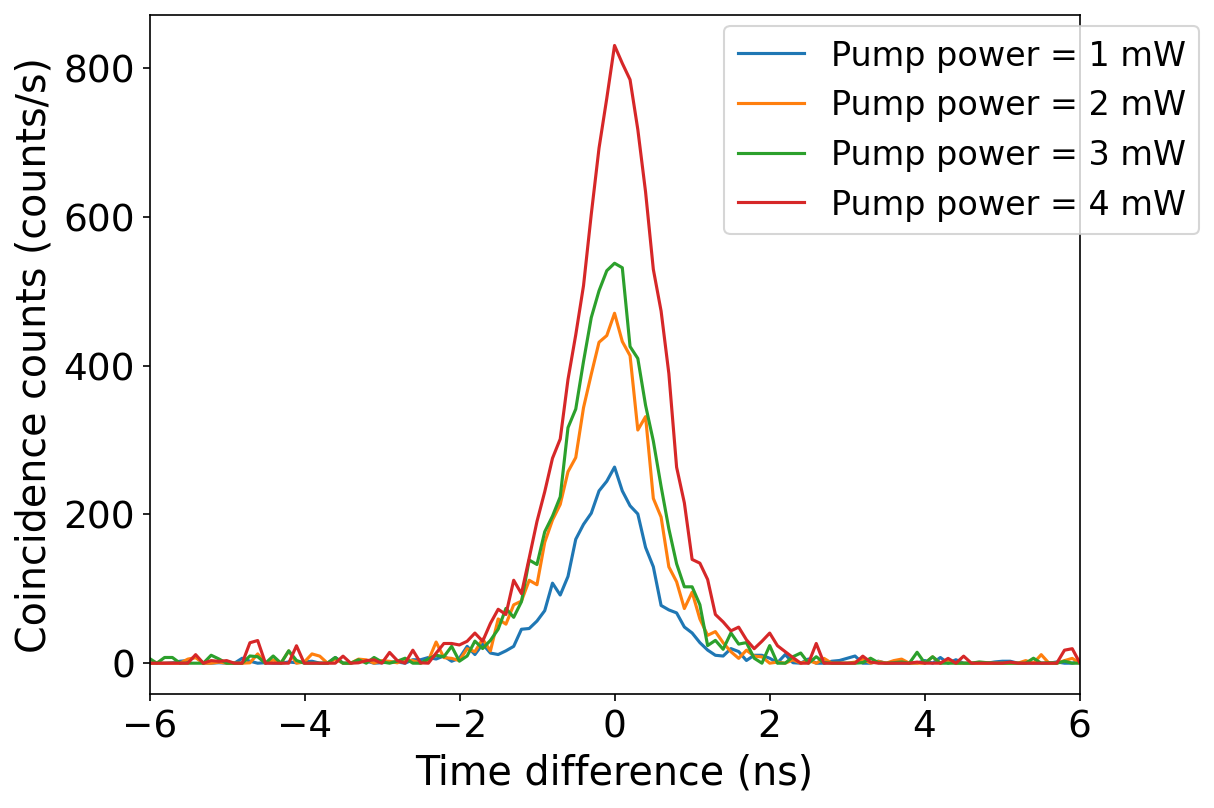}
\caption{Measured correlation histogram as a function of the delay between signal and idler arrival times after accidental counts subtraction. The coincidence peak width is $1.14\pm0.04$ ns.}
\label{fig:coincidence_peak}
\end{figure}

\begin{figure}[h]
    \centering
    \begin{subfigure}{0.535\textwidth}
        \includegraphics[width=\textwidth]{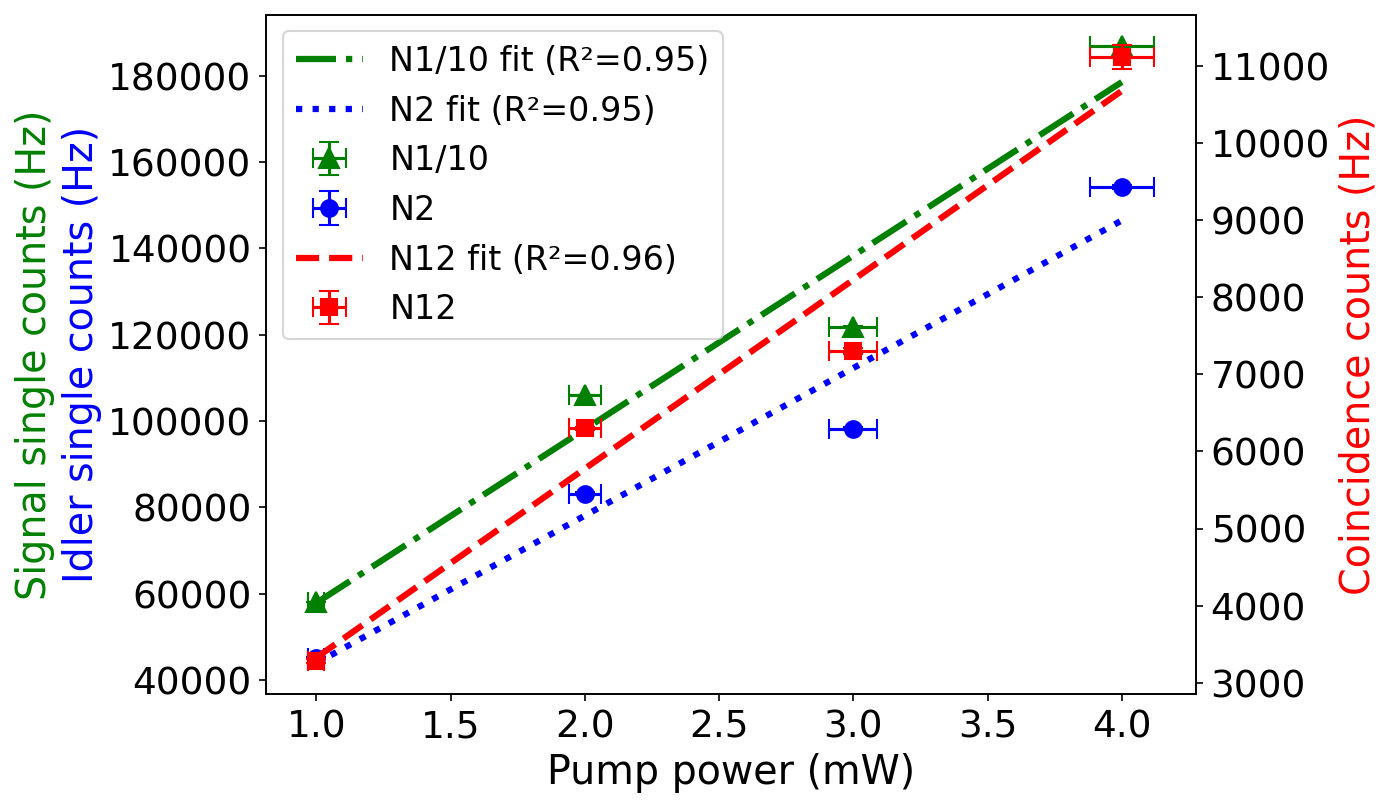}
        \label{fig:Coincidences_a}
    \end{subfigure}
		\hfill
    \begin{subfigure}{0.42\textwidth}
        \includegraphics[width=\textwidth]{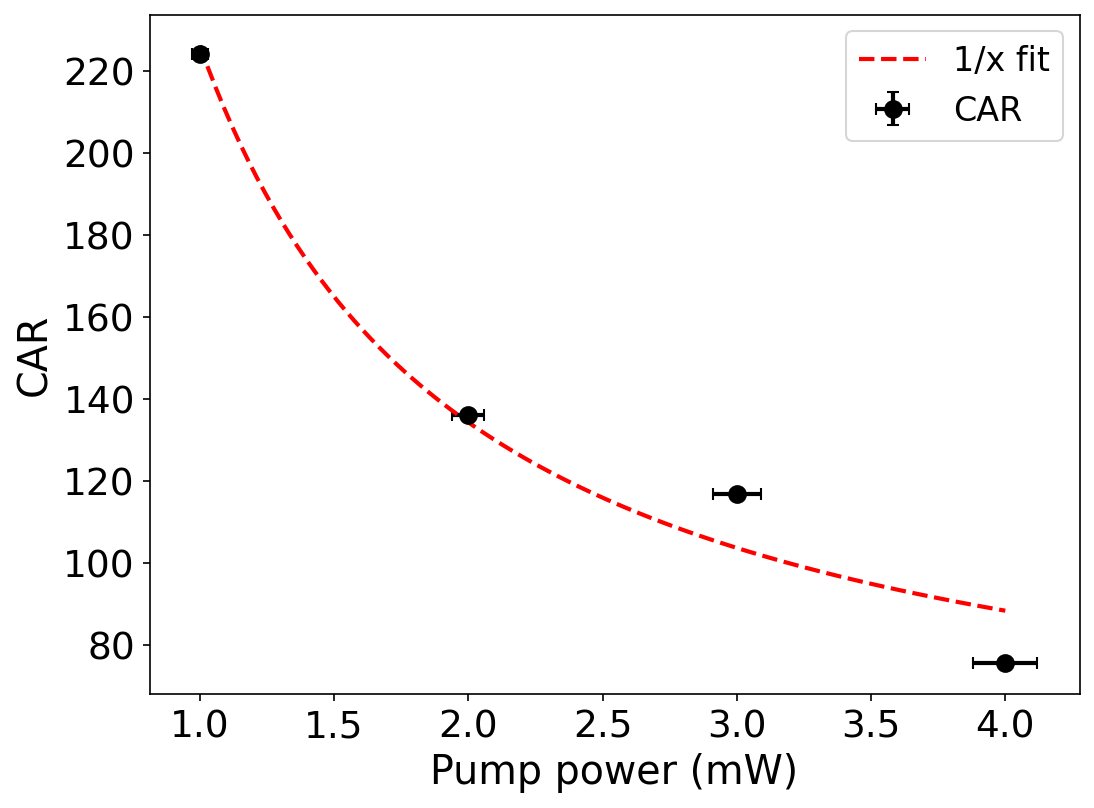}
        \label{fig:Coincidences_b}
    \end{subfigure}
    \caption{a) Measured single ($N_{1}/10$ and $N_{2}$) and coincidence counts ($N_{12}$) as a function of the pump power, after subtracting dark and accidental counts, respectively. To ease the graphical representation, the single counts $N_{1}$ are scaled down by a factor of ten. b) Measured coincidence to accidental ratio (CAR) as function of the pump power.}
    \label{fig:Coincidences}
\end{figure}

To demonstrate the generation of down-converted photon pairs, we measure the difference in arrival times between signal and idler photons detected at planes A and B in Fig.~\ref{fig:setup}, respectively, with SPCM-AQRH Excelitas single-photon detectors (SPDs) and a quTag correlation electronics from qutools (Fig.~\ref{fig:coincidence_peak}). Furthermore, we measure the effect of increasing the pump power on the UV-A SPDC source to demonstrate that it operates in the low gain regime (Fig.~\ref{fig:Coincidences}). To study the dependence of pump power on the SPDC source, pump power was varied using HWP1 in Fig.~\ref{fig:setup}, and further attenuated to a maximum of $4~\mathrm{mW}$ by placing a neutral density filter before the BBO crystal. The attenuation was necessary to prevent saturation of the SPDs and electronics, and avoid a non-linear response of the detectors. The detection range of the SPDs used spans from $\mathrm{400~nm}$ to $\mathrm{1060~nm}$ with peak sensitivity $65\%$ at $650$ nm. Therefore, to be able to detect idler photons, the BBO orientation was set to generate idler photons near $\mathrm{400~nm}$ instead of the shortest UV wavelength observed during the spectral analysis ($\mathrm{347.39~nm}$). In addition, a long pass filter with a cut edge at $\mathrm{355~nm}$ is placed at position B (see Fig.~\ref{fig:setup}) to block any residual pump light from reaching the detector for the idler beam. The dichroic mirror DM2 acts as a long pass filter for the signal beam and no additional filter is needed. 

The measured coincidence peaks (see Fig.~\ref{fig:coincidence_peak}) confirm the generation of correlated photon pairs with non-degenerate wavelengths for different pump powers. The peak is measured at zero time difference between signal and idler arrivals with a coincidence count rate of $>200~\mathrm{counts/s~mW}$ after accidental counts subtraction, and a peak width of $1.14\pm0.04$ ns.
%\mgb{how would the peak look if there would be more than just pairs generated? or is it only giving an idea of the time correlation?--> in the high gain, the main contribution still comes from 1 photon pair, the CAR is simply decreased because you have less of them and lots of multi-photon pairs. This graph demonstrates that you generate time-correlated photon pairs.} \mgb{compare with Arahata2021 and it's reference, Suezawa, with the pairs generation rate of $10^5 /s mW$.} 
%The coincidence peak width is \mgb{ explain a bit (coincidence peak is directly the coherence length of the pair, or the detector properties need to be taken into account?)--> yes, peak width also dependent on detector jitter, but important here is that the peak increases linearly with the pump --> not really lineraly... should we comment?}.

To corroborate the observation of SPDC photon-pairs, we measure the dependency of single-photon and coincidence counts with increasing pump power, finding a good agreement with a linear fit of the experimental points (see Fig.~\ref{fig:Coincidences_a}); this confirms that we operate in the low-gain regime, and contributions from multi-pair emission remain negligible. From the single and coincidence counts observed, the pairs generation rate can be calculated as $N_{\mathrm{pair}} = \frac{N_1N_2}{N_{12}}$~\cite{Simon_Sergienko} giving a rate of $ N_{\mathrm{pair}}=6.48\times 10^{6}\mathrm{counts}~\mathrm{s}^{-1} ~\mathrm{mW}^{-1}$. %\mgb{exposure time=1s, coincidence window=2ns, integration time=5s. Which one is the correct to give a $N_{\mathrm{pair}}$ in $\mathrm{counts} ~\mathrm{s}^{-1}~\mathrm{mW}^{-1}$? --> what is the difference between exposure time and integration time? Anyways, the correct one should be the total time during which you accumulate coincidence counts}.
Finally, we evaluate the coincidence-to-accidental ratio (CAR) and plot it against the input the pump power (Fig.~\ref{fig:Coincidences_b}). We measure CAR values as high as 220 for pump powers of 1 mW, showing that our photon-pair rate is well above the accidental background. Additionally, the CAR variation as function of the pump power is in reasonable agreement with an inversely proportional fit, further corroborating operation in the spontaneous regime: in the low-gain regime of SPDC, true coincidences scale linearly with the pump, while in contrast accidentals, assuming they originate from uncorrelated detection events, can be modeled with Poissonian statistics as $N_\mathrm{acc} \sim N_1 N_2 \tau /T$, where $N_{1/2}$ represent signal/idler single-photon counts, $\tau$ is the bin width and $T$ is the histogram integration time. Hence, accidentals quadratically scale with the pump power, implying that the CAR scales as $~1/x$ in the low gain regime.

\section{Conclusion}

We demonstrated a highly non-degenerate photon-pair source based on the second-order nonlinear process of SPDC, able to generate UV-A photons down to 347.39 nm, and we exploited the birifringence of the employed BBO nonlinear crystal to control the emission wavelengths. Compared to previous demonstrations of UV photon-pair generation based on third-order SFWM \cite{LOH23}, we demonstrated a several orders of magnitude higher photon-pair rate, albeit being limited in terms of emission wavelength by the 266 nm pump laser. Our results open potentially impactful application-oriented quantum sensing implementations, especially in the biomedical field  \cite{LUN05,ALM16,FGA14}. Among potential applications, autofluorophores NADH and FAD exhibit ideal excitation wavelengths between in the UV-A \cite{BLA14}, and possess a high potential to capture the metabolic state of cells, opening new avenues to improve personalized rapid, real-time quantum-sensing-driven immune cell phenotyping.

\section{Acknowledgements}
The authors thank Gil Zimmermann for his valuable comments.
This work was supported as a Fraunhofer LIGHTHOUSE PROJECT (QUILT). We further acknowledge support from the European Union’s Horizon 2020 Research and Innovation Action under Grant Agreement No. 304 101113901 (Qu-Test, HORIZON-CL4-2022-QUANTUM-05-SGA), and from the Free State of Thuringia under the funding ID 2021 FGI 0043.

%%%%%%%%%% If using BibTeX:
%\section{References}
%\normalem
\bibliography{UV-IR_Source_v1}

\end{document}